\documentstyle[10pt,epsfig]{article}
\def\bfl{\begin{flushleft}}
\def\efl{\end{flushleft}}
\def\bfr{\begin{flushright}}
\def\efr{\end{flushright}}
\def\bc{\begin{center}}
\def\ec{\end{center}}
\def\be{\begin{equation}}
\def\ee{\end{equation}}
\def\ba{\begin{eqnarray}}
\def\ea{\end{eqnarray}}
\def\nn{\nonumber }
\def\lb#1{\label{#1} }

\def\text#1{\mbox{#1} }
\def\drm{\text{d}}
\def\tableline{\hline}

\def\Sign#1{\, \text{sign}\left(#1\right) }
\def\Der#1#2{\,\frac{\partial #1}{\partial #2}}

\def\Kummer#1#2#3{\, \text{M}\left(#1,\,#2;\,#3 \right) } 

\def\Arccosh#1{\, \text{arccosh}\, #1}

\begin{document}

~\\
{\it General Relativity and Gravitation, Vol. 31, No. 12 (December), 
pp. 1821-1836, 1999}
\footnote{\normalsize \it
The late LANL e-version is slightly extended
with respect to that published in GRG (some complementary
speculations, footnotes and suggested references were added).}
~\\
~\\
\bc
{\LARGE \bf
Evolution of thin-wall configurations of texture matter
}

~~\\
{\large Konstantin G. Zloshchastiev}\\
~\\
E-Mail:  zlosh@email.com;
URL(s): http://zloshchastiev.webjump.com, http://zloshchastiev.cjb.net\\
~~\\

{\it
Received: January 15, 1999 (GRG), 
January 1, 2000 (LANL e-archive)}\\
\ec
~\\
\abstract{\normalsize
~\\
We consider the free 
matter of global textures within the framework of the
perfect fluid approximation in general relativity.
We examine thermodynamical properties of 
texture matter in comparison with radiation fluid and bubble matter.
Then we study dynamics of thin-wall selfgravitating texture objects, and
show that classical motion can be elliptical (finite), 
parabolical or hyperbolical.
It is shown that total gravitational mass of 
neutral textures in equilibrium equals to zero as was expected.
Finally, we perform the
Wheeler-DeWitt's minisuperspace quantization of the theory,
obtain exact wave functions and discrete spectra of bound states with
provision for spatial topology.
}

~\\

PACS number(s):  04.40.Nr, 04.60.Ds, 11.27.+d

~~\\

%\newpage
\large

\section{Introduction}\lb{sec:intro}

The scalar field theories, in which the global symmetry $G$ is spontaneously
broken to $H$ in such a way that vacuum manifold 
$G/H$ has nontrivial homotopy group $\pi_3 (G/H)$,
predict the existence of the 
matter with an equation of state $\varepsilon + 3 p = 0$
called as the texture matter or k-matter.\footnote{\normalsize
Also, the matter with
such an equation of state naturally appears in string cosmology theories.
One can obtain it from the ``string-driven'' 
effective equation of state $\varepsilon + d p - n q = 0$, 
where $d$ and $n$
are respectively the numbers of expanding and internal 
contracting dimensions,
$p$ and $q$ are respectively the pressure in the expanding space and 
shrinking dimensions \cite{gsv}.
Indeed, in the ``after-string'' era 
the internal dimensions were compactified (therefore, $n \to 0$),
the number
of expanding dimensions became the usual one, $d=3$.
This texture-dominated era continued till epoches of 
standard-model particles, quarks and leptons, 
when the decreasing temperature and 
density led textures to couple up. 
Nowadays we probably could observe some tracks of it not only at cosmological scales
but also among the fundamental properties of recently observed
particles \cite{zlo005}.
}
Of course, the terminology "texture matter" does not seem to be perfectly 
apposite because the real $\sigma$-model textures are dynamical defects
and the equation of state above is not valid in general case.
However, in numerous papers such a terminology was fixedly  
settled (see \cite{dn} and references therein) thus
we will follow it in present paper as well.\footnote{\normalsize
The dualities found in some 
string cosmology models suggest
that the matter with this equation of state
can be regarded as the matter of low-energy string origin 
which in some sense is
dual to radiation fluid (incoherent radiation).
Therefore, the term ``dual-radiation matter'' seems to be appropriate
for it as well.
}
Some known properties of textures say that it is probably another kind 
of vacuum similar to the de Sitter vacuum $\varepsilon + p = 0$
(the latter is known also as the bubble matter).
Let us consider, for instance, the O(4)$\to$O(3) textures arising in the
scalar fourplet theory
described by the action
\[
S (\vec\phi) = \int
\left[
       \partial^\mu \vec\phi\, \partial_\mu \vec\phi + \lambda
(\vec\phi\cdot\vec\phi - \eta^2)^2
\right]
 \sqrt{-g}\, \drm^4 x
\]
in a closed FRW universe ($0\leq\xi\leq\pi$)
\[
\drm s^2 = \drm t^2 - a^2(t) [\drm \xi^2 + \text{sin}^2 \xi
(\drm \theta^2 + \text{sin}^2\theta \drm \varphi^2)].
\]
Then the texture solution of winding number one,
\[
\vec\phi = \eta 
\left[
\begin{array}{r}
\cos{\varphi}\sin{\theta}\sin{\xi}\\
\sin{\varphi}\sin{\theta}\sin{\xi}\\
\cos{\theta}\sin{\xi}\\
\cos{\xi}
\end{array}
\right],
\]
has the following stress-energy tensor,
\[
T^\mu_\nu = \frac{\eta^2}{2 a^2}\, \text{diag} (3,1,1,1),
\]
which evidently satisfies with the above-mentioned equation of state.
The zero-zero component of this tensor
will be compared in Sec. 3 with a surface case.

The gravitational effects caused by 3D texture matter were 
intensively
studied in many works \cite{dn,texture}.
The main aim of present paper is to study the 2D fluid of global 
textures which forms spherically symmetric 
singular hypersurfaces (surfaces of
discontinuities of second kind).
These hypersurfaces can be interpreted both as the
thin-wall approximation of the layer of bulk matter and 
as the brane-like objects embedded in spacetime of higher dimensionality.
As such, the singular model turns to be simple enough to obtain
important and instructive exact results not only when studying 
classical dynamics but also when considering quantum aspects.
With respect to the 3D case this model appears to be the thin-wall
approximation, which can elicit main features common for 2D and 3D
cases.
 
The paper is organized as follows.
In section 2 we give a comparative description of thermodynamics
of 2D and 3D texture matter at finite temperature 
with respect to each other and
with respect to bubble matter and 
ordinary matter represented by radiation fluid.
Section 3 is devoted to classical dynamics of the isentropic
singular shells ``made'' from 2D texture fluid.
In section 4 we perform minisuperspace quantization of the singular
model with provision for both the through (wormhole-like)
and ordinary topology.
Conclusions are made in section 5.

\section{Comparative thermodynamics}

Let us consider the thermodynamical properties of texture matter as such
and in comparison with those for radiation fluid (quasi-counterpart of
texture) and bubble matter $\varepsilon + p = 0$.
First of all, we try to answer the question, what is thermodynamical 
information we can obtain from an equation of state.

The first thermodynamical law says:
\be                                                      \lb{eq1}
\drm E = T \drm S - p \drm V.
\ee
On the other hand, following the definition of the entropy as a function
of volume and temperature, one can write
\be                                                      \lb{eq2}
\drm S = \Der{S}{T} \drm T + \Der{S}{V} \drm V.
\ee
Comparing these equations, we obtain
\ba
&&\Der{S}{T} = \frac{1}{T} \Der{E}{T},                     \lb{eq3}\\
&&\Der{S}{V} = \frac{1}{T} 
             \left(
                   p + \Der{E}{V}
             \right).                                    \lb{eq4}
\ea
Then the equality of mixed derivatives yields the expression
\be                                                      \lb{eq5}
p + \Der{E}{V} = T \Der{p}{T},
\ee
which gives opportunities to obtain internal energy as a function
of volume and temperature from an equation of state.
Let us introduce the densities of energy and entropy such that
\be                                                      \lb{eq6}
E=\varepsilon (T) V,~~S=s(T) V,
\ee
and consider barotropic matter with linear equation of state (LEOS)
\be                                                      \lb{eq7}
p = \eta \varepsilon.
\ee
Then (\ref{eq5}) reads
\be                                                      \lb{eq8}
\eta T \frac{\drm \varepsilon}{\drm T} = (\eta +1) \varepsilon,
\ee
and we obtain the energy density
\be                                                      \lb{eq9}
\varepsilon = \varepsilon_0 T^{1+1/\eta}.
\ee
For instance, for 3D radiation fluid this expression yields
the expected Stefan-Boltzmann law describing energy of incoherent 
radiation with respect to temperature:
\[
\varepsilon = \alpha_{SB} T^4.
\]
The internal energy and pressure are, respectively,
\ba
&&E = \varepsilon_0 T^{1+1/\eta} V,                           \lb{eq10}\\
&&p = \eta \varepsilon_0 T^{1+1/\eta}.                        \lb{eq11}
\ea
Further, from (\ref{eq4}), (\ref{eq6}), 
(\ref{eq7}) and (\ref{eq9}) one can see that entropy has to be
\be                                                        \lb{eq12}
S = (\eta +1) \varepsilon_0 T^{1/\eta} V + S_0.
\ee
The above-mentioned special cases of LEOS matter are illustrated
in table \ref{tab-1}.\footnote{\normalsize
The features presented there (first of all, the 
unusual inverse dependence of energy and entropy on temperature
which means that increasing of temperature is energetically favorable)
can serve also
for revealing of free texture matter in present era, e.g., 
inside superhot objects 
with appropriate energy density 
(of order of GUT scale, $10^{13}$ TeV).}
We can observe, e.g., that texture matter
cannot approach zero temperature even formally (without the third law of
thermodynamics) because its energy diverges;
bubble and texture matter have nonzero minimal energy 
unlike ordinary matter $\eta > 0$ including ultrarelativistic radiation
fluid.
It seems to be another argument to the advantage of interpretation
of the texture matter as a specific vacuum state similar to the
de Sitter one.

\begin{table}
\caption{
Comparative thermodynamical properties of ordinary and vacuum-like matter.}
\bc
\begin{tabular}{llccl}
 \tableline
 Matter&EOS&Energy density&Entropy&Comments\\
 \tableline
 3D radiation fliud&$\varepsilon-3p=0$&$\varepsilon_0 T^4$&$\frac{4}{3} \varepsilon_0 T^3 V$ &Stefan-Boltzmann law\\
 2D radiation fliud&$\varepsilon-2p=0$&$\varepsilon_0 T^3$&$\frac{3}{2} \varepsilon_0 T^2 V$ &\\
 de Sitter bubble &$\varepsilon+p=0$&$\varepsilon_0$&$S_0$ &no dependence on $T$\\
 2D texture&$\varepsilon+2p=0$&$\varepsilon_0 T^{-1}$&$\frac{1}{2} \varepsilon_0 T^{-2} V + S_0$ &$T \not= 0$\\
 3D texture&$\varepsilon+3p=0$&$\varepsilon_0 T^{-2}$&$\frac{2}{3} \varepsilon_0 T^{-3} V + S_0$ &$T \not= 0$\\
 \tableline
 \end{tabular}
\ec
 \label{tab-1}
 \end{table}

\section{Thin-wall model}

Beginning from the classical works \cite{lan,dau,isr} 
formalism of surface layers has been widely described 
in the literature (see Refs. \cite{mtw,bb} for details).
The three-dimensional singular embeddings appear to be both 
interesting extended objects as such,
and simple (but realistic) models of four-dimensional phenomena.
From the viewpoint of the general physics of extended objects the concept 
``singular hypersurface'' has to be the next-order 
approximation, after the ``point particle'' one, 
which takes into account
both external, kinetic and dynamical, properties and
internal structure (surface pressure, mass density, 
temperature etc.).

So, one considers the infinitely thin isentropic layer of matter 
with the surface 
stress-energy tensor of a perfect fluid in general case 
(we use the units $\gamma=c=1$, where 
$\gamma$ is the gravitational constant)
\[
S_{ab}=\sigma u_a u_b + p (u_a u_b +~ ^{(3)}\!g_{ab}),
\]
where $\sigma$ and $p$ are the surface mass-energy density  and  pressure 
respectively, {\bf u} is the unit tangent vector, 
$~^{(3)}g_{ab}$ is the three-metric of the shell's hypersurface.
We suppose that this shell is spherically symmetric, closed, and 
hence divides 
the whole manifold into the two regions $\Sigma^\pm$.
Also we suppose the metrics of the space-times outside $\Sigma^+$ and inside 
$\Sigma^-$ of a spherically symmetric shell to be of the form
\be
\drm s_\pm^2 =
- [1+\Phi^{\pm}(r)] \drm t^2_\pm + [1+\Phi^{\pm}(r)]^{-1} \drm 
r^2 + r^2 \drm \Omega^2,                                      \lb{eq13}
\ee
where $\drm\Omega^2$ is the metric of the unit two-sphere.
Of course, we have some loss of generality but it is enough for further.
It is possible to show that if one introduces the proper time  
$\tau$, then the 3-metric of a shell can be written in the form
\be
^{(3)}\!\drm s^2 = - \drm \tau^2 + R^2 \drm \Omega^2,         \lb{eq14}
\ee
where $R(\tau)$ is a proper radius of a shell.
Define a simple jump of the second fundamental forms across a shell as 
$[K^a_b]=K^{a+}_b - K^{a-}_b$, where 
\be                                                           \lb{eq15}
K^{a\pm}_b = 
            \lim\limits_{n\to \pm 0} 
            \frac{1}{2} ~^{(3)}\!g^{a c} 
            \Der{}{n} ~^{(3)}\!g_{c b},
\ee
where $n$ is a proper distance in normal direction.
The Einstein equations on a shell then yield
equations which are the well-known Lichnerowicz-Darmois-Israel 
junction conditions 
\be
(K^a_b)^+ - (K^a_b)^- = 4 \pi\sigma (2 u^a u_b + \delta^a_b),  \lb{eq16}
\ee
Besides, an integrability condition of the Einstein equations is
the energy conservation law for shell matter.
In terms of the proper time it can be written as
\be
\drm \left( \sigma ~^{(3)}\!g \right) +
p~ \drm \left( ~^{(3)}\!g \right) 
+ ~^{(3)}\!g~   [ T ]\, \drm \tau =0,                        \label{eq17}
\ee
where $[ T ] = (T^{\tau n})^+ - (T^{\tau n})^-$,  
$T^{\tau n} = T_\alpha^\beta u^\alpha n_\beta$ is the
projection of stress-energy tensors in the $\Sigma^\pm$
space-times on the tangent and normal vectors, 
$^{(3)}\!g=\sqrt{-\det{(^{(3)}\!g_{ab})}} = R^2 \sin{\theta}$.

We assume that our shell carries no charges on a surface and contains
no matter inside itself.
If we define $M$ to be the total mass-energy of the shell then one can
suppose the external and internal spacetimes to be Schwarzschild and
Minkowskian respectively: 
\be
\Phi^+ = - \frac{2 M}{R},~~
\Phi^- = 0.                                               \label{eq18}
\ee
After straightforward computation of extrinsic curvatures the 
$\theta\theta$ component of (\ref{eq16}) yields the equation of motion
of the perfect fluid neutral hollow shell
\be
\epsilon_+ \sqrt{1+\dot R^2 - \frac{2 M}{R}} - 
\epsilon_- \sqrt{1+\dot R^2} = - \frac{m}{R},             \label{eq19}
\ee
where
\be
m = 4 \pi \sigma R^2                                     \label{eq20}
\ee
is interpreted as the (effective) rest mass, 
$\dot R=\drm R/\drm\tau$ is a proper velocity of the
shell, $\epsilon_+ = \Sign{\sqrt{1+\dot R^2 - 2 M/R }}$,
$\epsilon_- = \Sign{\sqrt{1+\dot R^2}}$.
It is well-known that $\epsilon = +1$ if $R$ increases in the outward
normal direction to the shell,
and $\epsilon = -1$ if $R$ decreases.
Thus, under the choice $\epsilon_+ = \epsilon_-$  we 
have an ordinary (black hole type) shell, whereas at 
$\epsilon_+ = -\epsilon_-$  
we have the thin-shell traversable wormhole \cite{vis}.

Let us consider the conservation law (\ref{eq17}).
One can obtain that
$[ T ]$ is identically zero for the spacetimes (\ref{eq18}).
Further, if we assume the 2D texture equation of state of the shell's 
matter,
\be
\sigma + 2p=0,                                         \label{eq21}
\ee
then, solving the differential equation (\ref{eq17}) with respect 
to $\sigma$,
we obtain
\be
\sigma  = \frac{\alpha}{2\pi R},                       \label{eq22}
\ee
hence 
\be
m  = 2\alpha R,                                        \label{eq23}
\ee
where $\alpha$ is the dimensionless integration constant which can 
be determined via surface mass density (or pressure) at fixed $R$.
The surface energy density determined by (\ref{eq22}) appears to be
the 2D analogue of the cosmological $T^0_0$ component from the
Sec. \ref{sec:intro} if
one takes into account the reduction of dimensionality.
This is an expected result: from the viewpoint of the 2D observer
``living'' on the shell it seems for him to be the whole universe with
the scale factor $R$.
Thus, our 2D fluid model indeed not only considers the established
trace properties of the texture stress-energy tensor but also 
restores its components for the surface case.
In this connection the integration constant $\alpha$ obtains the
sense of the (squared) topological charge $\eta$.
The topological nature of the textures 
will brightly show itself at the end of
of this section when we will study the texture fluid singular layers
with the vanishing total gravitational mass-energy.

\begin{table}
\caption{
Classification of texture shells with respect to $\alpha$.
The abbreviations ``OS'' and ``WS'' 
mean the ordinary and wormhole shells, 
respectively; ``$\star$'' denotes 
the impossibility of junction, ``$\forall$'' says that junction is
possible at any $\Sign{M}$ (trivial case $\alpha = 0$ is missed).}
\bc
\begin{tabular}{ccccc}
 \tableline
 &\multicolumn{2}{c}{OS} &\multicolumn{2}{c}{WS}\\
 &${\epsilon_+=1\choose{\epsilon_-=1}}$ &${\epsilon_+=-1\choose{\epsilon_-=-1}}$ &${\epsilon_+=1\choose{\epsilon_-=-1}}$ &${\epsilon_+=-1\choose{\epsilon_-=1}}$\\
 \tableline
 $\alpha>0$ &$M > 0$&$M < 0$&$\star$ & $\forall$ \\
 $\alpha<0$ &$M < 0$&$M > 0$&$\forall$ & $\star$ \\
 \tableline
 \end{tabular}
\ec
 \label{tab-2}
 \end{table}

Equations (\ref{eq19}) and (\ref{eq23}) together  
with the choice of the signs $\epsilon_\pm$ 
completely determine the motion of the thin-wall texture.
In conventional general relativity it is usually supposed that masses 
are nonnegative.
However, keeping in mind possible wormhole and 
quantum extensions of the theory \cite{man}, 
we will not restrict ourselves by positive values and consider 
general case of arbitrary (real) masses.
Then forbidden and permitted signs of this values can be
determined from table \ref{tab-2}.
Let us find now the trajectories of 2D textures.
Integrating (\ref{eq19}) we obtain the transcendental equation 
for $R(\tau)$
\be                                       \lb{eq24}
\tau/M = J (R/M) - J (R_0/M),
\ee
where
\begin{eqnarray}
&&J(y) = 
\left\{
      \begin{array}{ll}
      \frac{1}{\alpha^2-1}
      \left\{
            \frac{1}{\alpha} \sqrt{Z_1} +
            \frac{1}{2\sqrt{1-\alpha^2}}
            \arcsin{Z_2}
      \right\},&\alpha^2 < 1,\\
      \pm\frac{1}{6} \sqrt{4 y +1} (2 y -1),&\alpha = \pm 1,\\
      \frac{1}{\alpha^2-1}
      \left\{
            \frac{1}{\alpha} \sqrt{Z_1} -
            \frac{1}{2\sqrt{\alpha^2-1}}
            \Arccosh{Z_2}
      \right\},&\alpha^2 > 1,
      \end{array}
\right.                                                              \nn\\
&&Z_1 = \alpha^2 (\alpha^2 - 1) y^2 + \alpha^2 y + 1/4,              \nn\\
&&Z_2 = 2\alpha(\alpha^2 - 1) y + \alpha.                            \nn
\end{eqnarray}
Thus, in dependence on the parameter $\alpha^2$ one can
distinguish elliptical, parabolical and hyperbolical trajectories.
Let us consider below the consistency conditions which yield 
permitted domains of $\alpha$ and $y=R/M$ for each from three cases
$\alpha^2$.

({\it a}) Hyperbolic trajectories ($\alpha^2 > 1$).\\
Following (\ref{eq24}) the next two conditions should be satisfied jointly:
\be                                              \lb{eq25}
Z_1 \geq 0,~~Z_2 \geq 0.
\ee
Define
\ba                                              
&&y_\pm = -\frac{1}{2\alpha} \frac{1}{\alpha\pm 1},  \lb{eq26}\\
&&\bar y = \frac{1}{2(1-\alpha^2)},                 \lb{eq27}
\ea
and consider the two subcases:\\
({\it a.1}) $\alpha < -1$.
Then $y_+ < \bar y < y_- < 0$ and
inequalities (\ref{eq25}) can be reduced respectively to
\[
\{ y \leq y_+ \} \cup \{ y \geq y_- \},~~
y \leq \bar y,
\]
that yields
\be                                              \lb{eq28}
y \leq y_+.
\ee
({\it a.2}) $\alpha > 1$.
Then $y_- < \bar y < y_+ < 0$ and
inequalities (\ref{eq25}) can be reduced respectively to
\[
\{ y \leq y_- \} \cup \{ y \geq y_+ \},~~
y \geq \bar y,
\]
that yields
\be                                              \lb{eq29}
y \geq y_+.
\ee
Thus, inequalities (\ref{eq28}) and (\ref{eq29}) determine permitted
regions $\{\alpha, R/M\}$ for hyperbolical trajectories.

({\it b}) Elliptic trajectories ($\alpha^2 < 1$).\\
In the same way as above we can obtain the next restrictions:
\be                                              \lb{eq30}
Z_1 \geq 0,~~-1 \leq Z_2 \leq 1,
\ee
and consider the two subcases:\\
({\it b.1}) $-1 < \alpha < 0$.
Then $y_+ > 0 $ and $y_- < 0$, and
inequalities (\ref{eq30}) read
\be                                              \lb{eq31}
y_- \leq y \leq y_+.
\ee
({\it b.2}) $0<\alpha < 1$.
Then $y_+ < 0 $ and $y_- > 0$, and
\be                                              \lb{eq32}
y_+ \leq y \leq y_-.
\ee

({\it c}) Parabolic trajectories ($\alpha^2 = 1$).\\
We obtain that $y$ should obey
\be                                              \lb{eq33}
y \geq -1/4.
\ee

\begin{figure}
\centerline{
\epsfysize=.9\textwidth
\epsfbox{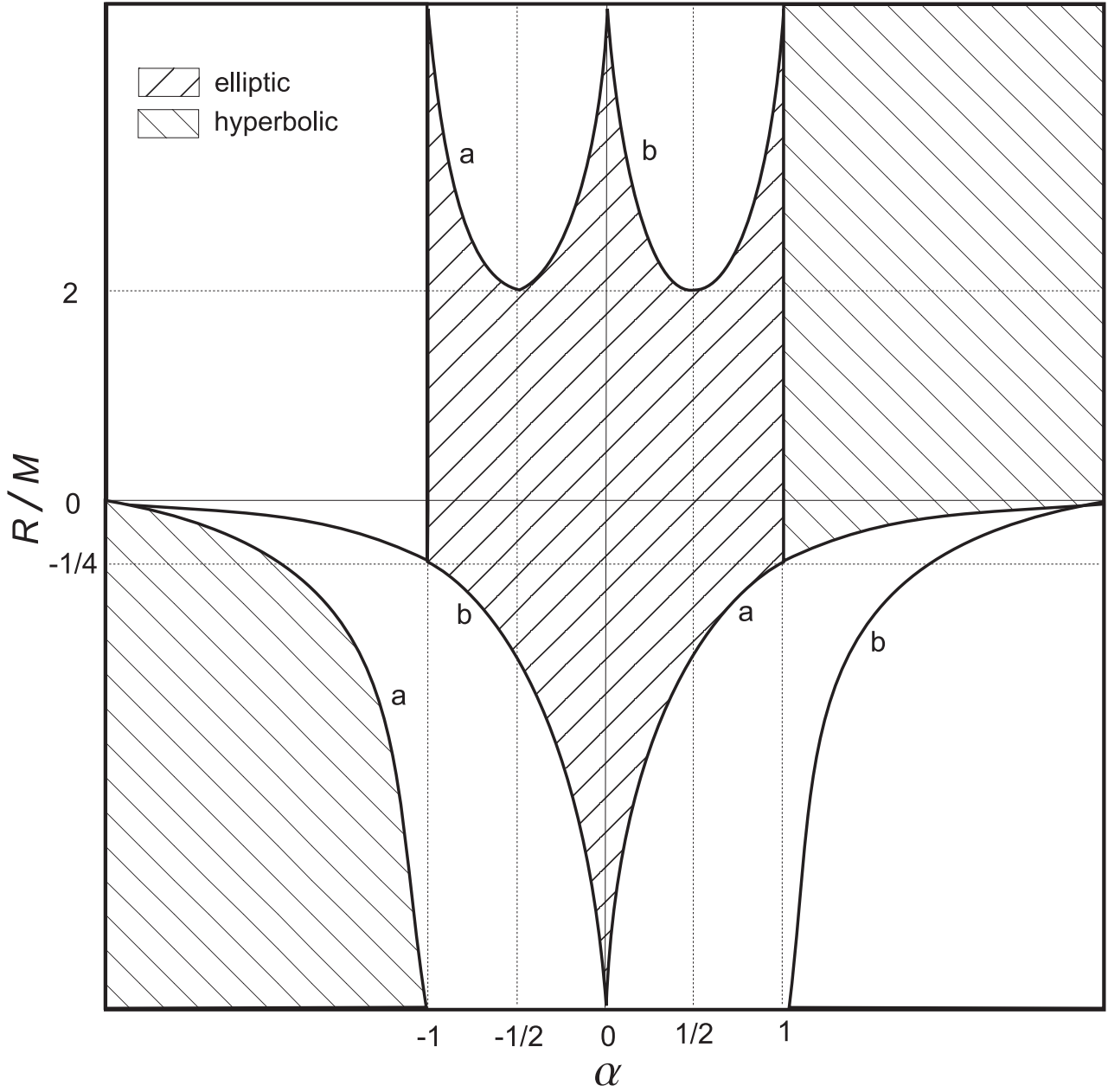}\\}
 \caption{Forbidden and permitted (dashed) regions of thin-wall texture 
motion. 
Permitted parabolical trajectories are the vertical half-lines 
$y \geq -/4$ at $\alpha=\pm 1$;
curve $a$ is $y=y_+$, $b$ is $y=y_-$.}
\label{fig1}
\end{figure}

The cases ({\it a})-({\it c}) are illustrated in figure \ref{fig1}
which represents dependence $y=R/M$ on $\alpha$.
Note, we did not restrict signs of mass,
therefore, table \ref{tab-2} should be kept in mind.

Let us study now equilibrium states of thin-wall textures.
Differentiating (\ref{eq19}) with respect to $\tau$, we obtain 
\be                                                \lb{eq34} 
\frac{\ddot R + M/R^2}{\epsilon_+ \sqrt{1+\dot R^2 - 2M/R}}-
\frac{\ddot R}{\epsilon_- \sqrt{1+\dot R^2}} = 0,
\ee
which independently could be obtained from 
junction conditions (\ref{eq16}).
Then in equilibrium state $\dot R = \ddot R = 0$ we obtain
\[                                                
M=0,
\]
i. e., the texture fluid in equilibrium 
has zero total gravitational mass that 
is already well-known \cite{dn}.
Another way to show this feature 
is to generalize (\ref{eq19}), (\ref{eq34})
by inserting the mass $M_-$ inside the shell, 
then the external and internal
spacetimes turn to be the Schwarzschild ones with masses $M^+$ and $M^-$
respectively.
Performing the analogical calculations we would obtain that in equilibrium:
at $\epsilon_+ =  \epsilon_-$ the static masses $M^+ = M^-$ and $\alpha=0$ 
(that evidently corresponds to the already decayed shell 
because $\alpha$ is the genuine criterion of existence and 
distinguishability of the shell)
whereas
at $\epsilon_+ = - \epsilon_-$ the static masses $M_\pm$ should vanish 
but $\alpha$ should not, giving the nonzero value for static radius,
i.e. again the static texture shell makes no contribution 
to the total gravitational mass of the system.

In other words, if in the (generalized) 
equations (\ref{eq19}), (\ref{eq34})
we even suppose $M^+ = M^- =0$ {\it identically} then we do not obtain
$\alpha \equiv 0$ with necessity.
It illustrates the fact that at some choice of signs $\epsilon_\pm$ 
we come to a non-trivial case despite the total masses are zero.
Indeed, at $\dot R = \ddot R = 0$ we have $\alpha \not= 0$ 
if $\epsilon_+ = - \epsilon_-$, i.e., for wormhole shells
(as for the ordinary hollow texture-shells, then always 
$\alpha_{\text{st}} = 0$,
and thus they cannot have equilibrium states).
Thus there exists the so-called zeroth traversable texture wormhole (ZTTW):
one can see that junction remains to be possible at 
$\epsilon_+ = - \epsilon_-$ and $\alpha \not= 0$ (among the rest linear
equations of state the texture's one 
(\ref{eq21}) appears to be unique in this sense).
Therefore, 
ZTTW represents itself the specific vacuum-like topological barrier 
(characterized only by $\alpha$, see eq. (\ref{eq22}) and 
comments after it) 
between two flat spacetimes\footnote{\normalsize It
does mean also that
in very general case the flat spacetime cannot be regarded as
{\it absolutely matter-free} one even on the classical level:
despite textures are the
defects of quantum-field nature and (after)string origin, 
their condensate, texture matter, is macroscopic. 
The locally flat spacetime is globally determined at least up to 
the foliation by nested ZTTW's walls.
}
which has no 
observable mass but possesses nontrivial internal structure and
inertial external dynamics
\ba
\sqrt{1+\dot R^2} = |\alpha|,\ \
\ddot R \equiv 0, \nn
\ea
thereby the restriction $\epsilon_- \alpha >0$ should be satisfied
as it can be readily seen from eqs.
(\ref{eq19}), (\ref{eq23}), (\ref{eq34}).

\section{Minisuperspace quantization}

Following the Wheeler-DeWitt's approach, in quantum cosmology the whole 
Universe is considered quantum mechanically and is described by a wave
function. 
The minisuperspace approach appears to be the direct application of
Wheeler-DeWitt's quantization procedure for (2+1)-dimensional singular
hypersurfaces having own internal three-metric 
(see \cite{vil,hkk,zlo001,zlo005} and references therein).
So, let us consider the minisuperspace model described by the 
Lagrangian:
\be                                                \lb{eq35}
L = \frac{m \dot R^2}{2} - \alpha (1-\alpha^2) R
+ \alpha M + \frac{M^2}{4\alpha R},
\ee
where $m$ was defined by (\ref{eq23}).
If we define
\[
U = \alpha (1-\alpha^2) R - \alpha M - \frac{M^2}{4\alpha R},
\]
then the equation of motion following from this Lagrangian is
\be                                                \lb{eq36}
\frac{\drm}{\drm\tau} (m \dot R) = 
\frac{m_{R} \dot R^2}{2} - U_{R},
\ee
where subscript ``$R$'' means derivative with respect to $R$.
Using time symmetry we can easy decrease order of this differential
equation and obtain
\be                                                \lb{eq37}
\dot R^2 = \frac{2}{\alpha R} (H-U),
\ee
where $H$ is the integration constant.
This equation coincides with double squared (\ref{eq19}) at (\ref{eq23})
when we suppose $H=0$ as a constraint.
Thus, our Lagrangian indeed describes dynamics of 
the thin-wall texture
up to the topological wormhole/blackhole division which was described by 
the signs $\epsilon_\pm$.
However, we always can restore the topology $\epsilon_\pm$
both at classical (rejecting redundant roots) and quantum 
(considering appropriate boundary conditions for the corresponding 
Wheeler-DeWitt equation, see below) levels.

Further, at $\Pi = m \dot R$ the (super)Hamiltonian is 
\be                                                \lb{eq38}
{\cal H} = \Pi \dot R - L = H = 0.
\ee
The prefix ``super'' means that in general
case ${\cal H}$ has to be a functional defined on the superspace 
which is the space of all admissible metrics and accompanying fields.
In spherically symmetric case the world sheet of a singular
hypersurface is determined by a single function, proper radius $R(\tau)$.

To perform quantization we replace momentum by the operator 
$\hat\Pi= -i \partial/\partial R$ \cite{vil} (we assume Planckian units), 
and
(\ref{eq38}) yields the Wheeler-DeWitt equation for
the wave function $\Psi (R)$:
\be                                                \lb{eq39}
\Psi_{RR} + 
\biggl[
      M^2 + 4 M \alpha^2 R - 4\alpha^2 (1-\alpha^2) R^2
\biggr] \Psi = 0.
\ee
One can see the main advantage of the minisuperspace approach, namely,
it does not require any time slicing on the basic manifold. 

Further, the important remark should be made now.
Last equation can be reduced to that for 
quantum harmonical oscillator, but not in all cases:
the oscillator's equation is defined on the 
line $(-\infty,+\infty)$ whereas
in our case the extension of an application domain on the whole
axis $R\in (-\infty,+\infty) $ seems to be physically ill-grounded
in the major cases, therefore 
we should study the quantum theory on the half-line $[0,\infty)$.
Strictly speaking, such a situation happens also in quantum field theory
then the mathematical procedure known
as the Langer modification had been used there \cite{raj}.
The similar transformation we perform below to obtain the required 
solution.

In the case $ R\in [0, +\infty) $ equation (\ref{eq39}) in general cannot
be resolved in terms of the parabolical cylinder functions and Hermite 
polynomials.
Fortunately, a solution can be expressed in terms of the 
functions well-defined on the half-line $[0, +\infty) $.
To show up this feature let us perform, at first, 
the following substitution
\be                                                \lb{eq44}
x = R - b/2a~ \Rightarrow ~ x\in [-b/2a, +\infty),
\ee
where $a=4\alpha^2 (1-\alpha^2)$, $b=4\alpha^2 M$.
Then (\ref{eq39}) can be rewritten as 
\be                                                \lb{eq45}
\Psi_{xx} + 
\frac{\Delta - 4 a^2 x^2}{4 a} \Psi = 0,
\ee
where $\Delta = (4\alpha M)^2$.
Further, considering (\ref{eq44}) it can be seen that at $a>0$ 
($\alpha^2 < 1$) the substitution
\be                                                \lb{eq46}
z = \sqrt{a} x^2
\ee
has not to be the injective mapping and acts like the baker's 
transformation \cite{moo} around $x=0$ that provides the
important property
\be                                                \lb{eq47}
z \in [0, +\infty).
\ee
Then the transformation
\be                                                \lb{eq48}
\Psi (x) = e^{-z/2} \sqrt{z} \omega (z)
\ee
rewrites (\ref{eq45}) in a form of the confluent hypergeometric equation
\be                                                \lb{eq49}
z \omega_{zz} + (c^\prime -z) \omega_z - a^\prime \omega =0,
\ee
whose solutions are the regular Kummer functions 
$\Kummer{a^\prime}{c^\prime}{z}$,
where
\[
a^\prime = \frac{3}{4} - \frac{\Delta}{16 a^{3/2}},~~
c^\prime = \frac{3}{2}.
\]
Therefore, the true solutions of (\ref{eq39}) at $ R\in [0, +\infty) $
are the functions (up to multiplicative constant):
\be                                                \lb{eq50}
\Psi = e^{-z/2} \sqrt{z} 
\Kummer{
        \frac{3}{4} - \frac{\Delta}{16 a^{3/2}}
       }
       {
        \frac{3}{2}
       }{z},
\ee
where
\be                                                \lb{eq51}
z = 2 |\alpha| \sqrt{1-\alpha^2}
\left[
R - \frac{M}{2 (1-\alpha^2)}
\right]^2.
\ee

Further, if we wish to determine bound states we should
require $\Psi (R=+\infty)=0$;
the also required condition  
$\Psi (R=0)=0$ (that corresponds, according to aforesaid, 
to $\Psi (z=0) = 0$) has been already satisfied
by the choice of the solution (\ref{eq50}) when a one 
integration constant was used 
(the second constant always remains to normalize a solution) 
\cite{zlo001,zlo005}.
In this case 
\be                                                 \lb{eq52}
a^\prime=-n,
\ee
$n$ is a nonnegative integer,
and the Kummer function moves to the Laguerre polynomials.
From last expression we obtain the mass spectrum 
of the thin-wall ordinary texture in a bound state:
\be                                                \lb{eq53}
M_n = \pm \sqrt{2|\alpha| (4 n + 3)} (1-\alpha^2)^{3/4},
\ee
which evidently has to be a subset of the oscillator's spectrum,
the Laguerre polynomials are connected with the Hermite ones
through the transformation (\ref{eq46}).
Thus, our procedure has cut out from
the oscillator's eigenfunctions and eigenvalues
those which satisfy with the boundary conditions on a half-line.

\section{Conclusion}

In present paper we considered texture matter and singular hypersurfaces
made from it.
First of all, we studied thermodynamical properties of 2D and 3D
texture matter in comparison with radiation fluid and bubble matter.
These properties say that textures can be imagined as the specific
vacuum state having the congeniality with the already known de Sitter 
vacuum and duality with radiation fluid \cite{gsv}.
We obtained equations of motion of selfgravitating texture objects,
showed that classical 
motion can be elliptical (finite), parabolical or hyperbolical,
thereby permitted and forbidden regions of motion was determined.
We showed up that neutral textures in equilibrium have
zero total gravitational mass as was expected.
Moreover, it was established that there can exist 
the nontrivial wormhole-textures
having vanishing total mass and matching two flat spacetimes.
Finally, we considered quantum aspects of the theory by means of
Wheeler-DeWitt's minisuperspace quantization procedure,
obtained the exact wave function and spectrum of bound states.

\def\ApPh {Astropart. Phys.}
\def\CJP  {Czech. J. Phys. }
\def\CQG  {Class. Quantum Grav.}
\def\EPL  {Europhys. Lett.}
\def\GRG  {Gen. Rel. Grav.}
\def\IJMP {Int. J. Mod. Phys.}
\def\IJTP {Int. J. Theor. Phys.}
\def\MPL  {Mod. Phys. Lett.}
\def\NPh  {Nucl. Phys.}
\def\PhE  {Phys. Essays}
\def\PhL  {Phys. Lett.}
\def\PhR  {Phys. Rev.}
\def\PhRL {Phys. Rev. Lett.}
\def\PhRp {Phys. Rep.}
\def\NCim {Nuovo Cim.}
\def\NuPh {Nucl. Phys.}
\def\prp {report}
\def\Prp {Preprint}

\def\jn#1#2#3#4#5{{#1}{#2} {\bf #3}, {#4} {(#5)}}
% #1 tittle  #2 ser  #3 vol  #4 page  #5 year
\def\boo#1#2#3#4#5{{\it #1} ({#2}, {#3}, {#4}){#5}}
% #1 tittle  #2 publisher  #3 place  #4 year  #5 page/, p.789/
\def\prpr#1#2#3#4#5{{ #3}{#4}{ (#5)}}                 %IOP style
% #1 tittle  #2 place #3 \Prp #4No #5 year 

%\newpage

%\newpage
%============< FIGURE >==============%
%======================================%

\end{document}